\documentclass[a4paper,fleqn,usenatbib,onecolumn]{mnras}

\usepackage{newtxtext,newtxmath}
\usepackage[T1]{fontenc}
\usepackage{ae,aecompl}
\usepackage{graphicx}
\usepackage{amsmath}
\usepackage{amssymb}

\newcommand\vx{\mathbfit{x}}
\newcommand\vk{\mathbfit{k}}
\newcommand\vs{\mathbfit{s}}
\newcommand\hP{\hat{P}}
\newcommand\hxi{\hat{\xi}}
\newcommand\hx{\hat{\mathbfit{x}}}
\newcommand\hk{\hat{\mathbfit{k}}}
\newcommand\hs{\hat{\mathbfit{s}}}
\newcommand\hz{\hat{\mathbfit{0}}}

\title[6dFGS multipoles]{Power spectrum multipoles on the curved sky:
  an application to the 6-degree Field Galaxy Survey}

\author[Blake et al.]{Chris Blake,$^1$\thanks{E-mail:
    cblake@swin.edu.au} Paul Carter$^2$ and Jun Koda$^{3,4}$ \\ $^1$
  Centre for Astrophysics \& Supercomputing, Swinburne University of
  Technology, P.O.\ Box 218, Hawthorn, VIC 3122, Australia \\ $^2$
  Institute of Cosmology \& Gravitation, University of Portsmouth,
  Dennis Sciama Building, Portsmouth, PO1 3FX, U.K. \\ $^3$
  Dipartimento di Matematica e Fisica, Universit\`{a} degli Studi Roma
  Tre, via della Vasca Navale 84, Rome 00146, Italy \\ $^4$ INFN
  Sezione di Roma 3, via della Vasca Navale 84, Rome 00146, Italy}
 
\date{Accepted XXX. Received YYY; in original form ZZZ}

\pubyear{2017}

\begin{document}
\label{firstpage}
\pagerange{\pageref{firstpage}--\pageref{lastpage}}
\maketitle

\begin{abstract}
The peculiar velocities of galaxies cause their redshift-space
clustering to depend on the angle to the line of sight, providing a
key test of gravitational physics on cosmological scales.  These
effects may be described using a multipole expansion of the clustering
measurements.  Focussing on Fourier-space statistics, we present a new
analysis of the effect of the survey window function, and the
variation of the line of sight across a survey, on the modelling of
power spectrum multipoles.  We determine the joint covariance of the
Fourier-space multipoles in a Gaussian approximation, and indicate how
these techniques may be extended to studies of overlapping galaxy
populations via multipole cross-power spectra.  We apply our
methodology to one of the widest-area galaxy redshift surveys
currently available, the 6-degree Field Galaxy Survey, deducing a
normalized growth rate $f \sigma_8 (z = 0.06) = 0.38 \pm 0.12$ in the
low-redshift Universe, in agreement with previous analyses of this
dataset using different techniques.  Our framework should be useful
for processing future wide-angle galaxy redshift surveys.
\end{abstract}


\section{Introduction}

The velocities of galaxies within the expanding Universe are generated
from underlying density fluctuations by gravitational physics,
producing the overall growth of cosmic structure with time.  The
statistics of these velocities are observable through the correlated
Doppler shifts they induce in the galaxy redshifts measured by a
spectroscopic survey, known as Redshift-Space Distortions (RSD).  This
effect imprints an anisotropy in redshift-space galaxy clustering with
respect to the local line of sight, which may be used to perform
precise tests of gravity on cosmological scales.

The measurement of RSD has become a standard application of modern
galaxy redshift surveys
\citep[e.g.,][]{Blake11,Howlett15,Alam17,Pezzotta17}, which has
allowed the growth rate of structure to be measured with approximately
$10\%$ accuracy across the redshift range $z < 1$.  Standard
treatments of RSD clustering statistics exist in both configuration
space and Fourier space, featuring various advantages and
disadvantages in terms of systematic modelling errors, statistical
signal-to-noise and algorithm performance.  In this study we develop
several new results regarding the application of Fourier-space
statistics.

The clustering anisotropy induced by galaxy velocities may be
conveniently described by a multipole expansion of the clustering
statistics with respect to the local line of sight
\citep{Cole94,Hamilton98}, permitting a powerful compression of the
information.  The variation of the line-of-sight direction across the
survey volume -- or ``curved-sky'' effect -- creates complications for
algorithms that evaluate Fourier-space clustering statistics using
Fast Fourier Transforms (FFTs) \citep{Yamamoto06}.  A series of
studies have developed techniques to apply power spectrum multipoles
addressing some of the difficulties created by the curved sky
\citep{Beutler14,Bianchi15,Scoccimarro15,Slepian15a,Slepian16,GilMarin16,Wilson17,Beutler17,Hand17,Castorina18,Sugiyama18}.
These studies explore estimating multipole power spectra including the
effect of the curved sky, modelling multipole power spectra in the
presence of a survey window function, applying these algorithms to
galaxy surveys (including the suppression of systematics in some
modes), and extending these treatments to higher-order statistics such
as the bispectrum.

We extend these results in three main areas.  First, we present a new
technique for evaluating model multipole power spectra in the presence
of a survey window function and curved sky.  Our method, which is
computed purely in Fourier space using FFT-based techniques, is an
alternative formulation of the mathematics presented by
\citet{Wilson17} and \citet{Beutler17}.  Second, we estimate the joint
covariance of power spectrum multipoles using an analytical Gaussian
approximation, including window function and curved-sky effects.  The
covariance of clustering statistics is often determined by applying
estimators to a large ensemble of mock catalogues, which are built to
match the galaxy survey properties as closely as possible.  Whilst
this approach enables the inclusion of relevant non-linear effects,
realistic mocks may sometimes be difficult to produce in sufficient
numbers, causing difficulties in evaluating the likelihood
\citep[e.g.,][]{Hartlap07,Taylor14}.  Analytical approaches to the
covariance are therefore also valuable
\citep[e.g.,][]{FKP,Xu12,Slepian15b,OConnell16,Grieb16,Mohammed17,Hand17,Howlett17}.
We provide new results that extend the calculations of \citet{FKP} in
order to determine the covariance of power spectrum multipoles in the
Gaussian approximation, including window function and curved-sky
effects.  Finally, a number of authors have pointed out that a joint
clustering analysis of overlapping galaxy populations, which share
correlated sample variance, improves the accuracy of measuring several
combinations of cosmological parameters
\citep[e.g.,][]{McDonald09,GilMarin10,Abramo12,Blake13}.  We extend
our convolution and covariance calculations to include the multipole
cross-power spectra of overlapping galaxy tracers.

We illustrate some of our new algorithms by performing a multipole
power spectrum analysis of one of the widest-area local galaxy
redshift surveys, the 6-degree Field Galaxy Survey
\citep[6dFGS,][]{Jones09}.  Configuration-space RSD studies of 6dFGS
have already been carried out by \citet{Beutler12} and
\citet{Achitouv17}; we present the first Fourier-space treatment.  Our
techniques should prove useful for the next generation of wide-area
spectroscopic studies such as the Taipan Galaxy Survey
\citep{daCunha17}, the Dark Energy Spectroscopic Instrument
\citep{DESI} and the 4MOST Cosmology Redshift Survey \citep{4MOST}.

Our paper is structured as follows.  In Section \ref{secform} we
introduce the power spectrum multipoles formalism, presenting new
calculations regarding the convolution with the survey window function
(Section \ref{secconv}) and the application to multiple galaxy tracers
(Section \ref{secmult}), and in Section \ref{seccov} we develop the
analytical covariance in the Gaussian approximation.  In Section
\ref{sec6dfgs} we present the analysis of 6dFGS data and mocks, and in
Section \ref{secsummary} we summarize the results.

\section{Power spectrum multipoles formalism}
\label{secform}

In this Section we summarize a purely Fourier-space scheme to analyze
the multipole power spectra of a galaxy redshift survey, including
curved-sky effects.  We characterize the survey by a window function
$\langle n(\vx) \rangle$, which predicts the galaxy number density as
a function of position in the absence of clustering (with the angle
brackets indicating an average over many realizations), and include a
position-dependent weight function $w(\vx)$ that may be used to
optimize the signal-to-noise of the measured statistics.

\subsection{Estimating the power spectrum multipoles}
\label{secest}

We first outline how power spectrum multipoles, $P_\ell(k)$, may be
estimated from a galaxy distribution.  The multipoles are defined by
an expansion of the anisotropic galaxy power spectrum $P(\vk)$, as a
function of wavevector $\vk$, with respect to a global line of sight:
\begin{equation} 
P(\vk) = \sum_\ell P_\ell(k) \, L_\ell(\mu) ,
\end{equation}
where $L_\ell$ are Legendre polynomials, and $\mu$ is the cosine of
the angle between $\vk$ and the line of sight (assuming azimuthal
symmetry).  However, the line-of-sight direction is not fixed, but
varies across the galaxy survey.  In the region of space around
position $\vx$, we can write $\mu = \hx.\hk$ such that for a local
line of sight,
\begin{equation}
  P(\vk,\vx) = \sum_\ell P_\ell(k) \, L_\ell(\hk.\hx) .
\label{eqpk}
\end{equation}
Inverting Equation \ref{eqpk} and averaging the statistic over all
positions, we can evaluate the power spectrum multipoles as
\begin{equation}
P_\ell(k) = \frac{2\ell+1}{V} \int d^3\vx \int \frac{d\Omega_k}{4\pi}
\, P(\vk,\vx) \, L_\ell(\hk.\hx) ,
\label{eqpkl1}
\end{equation}
where $d\Omega_k$ integrates over all angles $\hk$.  In this
formulation we are applying the ``local plane-parallel approximation''
defined by \citet{Beutler14}, which assumes that the position vectors
$(\vx,\vx')$ of a pair of galaxies separated by relevant scales are
locally parallel $(\hk.\hx \approx \hk.\hx')$, but we account for the
changing line-of-sight direction between different pairs.  This is a
good approximation for our statistics and scales of interest
\citep{Papai08,Samushia12,Yoo15}.  \citet{Castorina18} recently
presented a systematic exploration of wide-angle effects beyond the
local plane-parallel approximation.

We can connect Equation \ref{eqpkl1} to the distribution of galaxies
across space by writing the power spectrum (in standard $h^{-3}$
Mpc$^3$ volume units) as the Fourier transform of the 2-point galaxy
correlation function $\xi(\vx,\vx') = \langle \delta(\vx) \,
\delta(\vx+\vs) \rangle$, as a function of galaxy overdensity $\delta$
and vector separation $\vs = \vx' - \vx$, as
\begin{equation}
P(\vk,\vx) = \int d^3\vs \, \xi(\vx,\vx') \, e^{i\vk.\vs} .
\end{equation}
We obtain
\begin{equation}
P_\ell(k) = (2\ell+1) \int \frac{d\Omega_k}{4\pi} \, \frac{1}{V}
\int d^3\vx \int d^3\vx' \, \xi(\vx,\vx') \, e^{i\vk.(\vx-\vx')} \,
L_\ell(\hk.\hx') .
\label{eqpkl2}
\end{equation}
The form of Equation \ref{eqpkl2} provides the following estimator for
the power spectrum multipoles
\citep{Yamamoto06,Bianchi15,Scoccimarro15}:
\begin{equation}
\hP_\ell(k) = \frac{2\ell + 1}{I} \int \frac{d\Omega_k}{4\pi} \left[
  \frac{1}{V} \int d^3\vx \int d^3\vx' \, \delta_w(\vx) \,
  \delta_w(\vx') \, e^{i\vk.(\vx-\vx')} \, L_\ell(\hk.\hx') -
  N_\ell(\vk) \right] ,
\label{eqest}
\end{equation}
in terms of the weighted galaxy overdensity $\delta_w$ computed from
the galaxy density field $n(\vx)$,
\begin{equation}
  \begin{split}
    \delta_w(\vx) &= w(\vx) \left[ n(\vx) - \langle n(\vx) \rangle \right] \\
    &= w(\vx) \, n(\vx) - n_w(\vx) ,
  \end{split}
\end{equation}
where $n_w(\vx) = w(\vx) \langle n(\vx) \rangle$, the normalization
\begin{equation}
I = \frac{1}{V} \int d^3\vx \, n_w^2(\vx) ,
\end{equation}
and a $\vk$-dependent shot noise term that affects all
multipoles\footnote{We assume that the window function $\langle n(\vx)
  \rangle$ has been evaluated using a sufficient number of random
  objects that it does not contribute to the shot noise, i.e.\ we take
  the limit $\alpha \rightarrow 0$ in the notation of \citet{FKP}.}
\begin{equation}
  N_\ell(\vk) = \frac{1}{V} \int d^3\vx \, w^2(\vx) \, \langle n(\vx)
  \rangle \, L_\ell(\hk.\hx) .
\end{equation}
These equations agree with those presented in Section 2 of
\citet{Bianchi15} (see also \citealt{Scoccimarro15} and
\citealt{Hand17}).  We refer the reader to \citet{Bianchi15} for a
description of FFT-based methods for evaluating this estimator, which
we apply in our analysis.

\subsection{Evaluating the model power spectrum multipoles}
\label{secconv}

We now consider how power spectrum multipole model predictions may be
evaluated for comparison with these estimators, including the effects
of the survey window function and curved sky (in the local
plane-parallel approximation).  We aim to obtain an expression purely
in Fourier co-ordinates, which may be evaluated using FFTs.  We
evaluate the expectation value $\langle \hP_\ell(k) \rangle$ of
Equation \ref{eqest} using \citep[e.g.,][]{FKP}
\begin{equation}
\langle \delta_w(\vx) \, \delta_w(\vx') \rangle = w(\vx) \, w(\vx') \,
\langle n(\vx) \rangle \, \langle n(\vx') \rangle \, \xi(\vx,\vx') +
w^2(\vx) \, \langle n(\vx) \rangle \, \delta_D(\vx-\vx') / V ,
\label{eqdel}
\end{equation}
where $\delta_D(\vx)$ is the Dirac $\delta$-function, normalized such
that $(1/V) \int d^3\vx \, f(\vx) \, \delta_D(\vx-\vx_0) = f(\vx_0)$.
The first and second terms in Equation \ref{eqdel} represent the
contribution to the density covariance of sample variance and shot
noise, respectively.  Substituting Equation \ref{eqdel} in Equation
\ref{eqest} we find
\begin{equation}
  \langle \hP_\ell(k) \rangle = \frac{2\ell + 1}{I} \int
  \frac{d\Omega_k}{4\pi} \, \frac{1}{V} \int d^3\vx \, n_w(\vx) \int
  d^3\vx' \, n_w(\vx') \, \xi(\vx,\vx') \, e^{i\vk.(\vx-\vx')} \,
  L_\ell(\hk.\hx') .
  \label{eqpkl3}
\end{equation}
We relate the expectation value $\langle \hP_\ell(k) \rangle$ to the
underlying model power spectrum statistics using
\begin{equation}
\xi(\vx,\vx') = \int \frac{d^3\vk'}{(2\pi)^3} \, P(\vk',\vx') \,
e^{-i\vk'.(\vx-\vx')} ,
\label{eqxi}
\end{equation}
to obtain
\begin{equation}
  \langle \hP_\ell(k) \rangle = \frac{2\ell + 1}{I} \int
  \frac{d\Omega_k}{4\pi} \, \frac{1}{V} \int d^3\vx \, n_w(\vx) \int
  d^3\vx' \, n_w(\vx') \int \frac{d^3\vk'}{(2\pi)^3} \, P(\vk',\vx')
  \, e^{i(\vk-\vk').(\vx-\vx')} \, L_\ell(\hk.\hx') .
\end{equation}
Substituting in the multipole expansion of Equation \ref{eqpk} we can
write this relation in the form
\begin{equation}
  \begin{split}
    \langle \hP_\ell(k) \rangle &= \frac{2\ell + 1}{I} \int
    \frac{d\Omega_k}{4\pi} \, \frac{1}{V} \int d^3\vx \, n_w(\vx) \int
    d^3\vx' \, n_w(\vx') \int \frac{d^3\vk'}{(2\pi)^3} \, \sum_{\ell'}
    P_{\ell'}(k') \, L_\ell(\hk.\hx') \, L_{\ell'}(\hk'.\hx') \,
    e^{i(\vk-\vk').(\vx-\vx')} \\
    &= \frac{2\ell + 1}{I} \int \frac{d\Omega_k}{4\pi} \, \frac{1}{V}
    \int \frac{d^3\vk'}{(2\pi)^3} \, \sum_{\ell'} P_{\ell'}(k') \,
    \int d^3\vx \, n_w(\vx) \, e^{i(\vk-\vk').\vx} \int d^3\vx' \,
    n_w(\vx') \, L_\ell(\hk.\hx') \, L_{\ell'}(\hk'.\hx') \,
    e^{-i(\vk-\vk').\vx'} \\
    &= \frac{2\ell + 1}{I} \int \frac{d\Omega_k}{4\pi} \, \sum_{\ell'}
    \frac{V}{(2\pi)^3} \int d^3\vk' \, P_{\ell'}(k') \,
    \tilde{n}_w(\vk-\vk') \, S_{\ell,\ell'}^*(\vk,\vk') ,
  \end{split}
\label{eqpkl4}
\end{equation}
where
\begin{equation}
S_{\ell,\ell'}(\vk,\vk') = \frac{1}{V} \int d^3\vx \, n_w(\vx) \,
L_\ell(\hk.\hx) \, L_{\ell'}(\hk'.\hx) \, e^{i(\vk-\vk').\vx} .
\label{eqsl}
\end{equation}
Physically, $S_{\ell,\ell'}(\vk,\vk')$ describes the effect of the
window function in causing a measurement of the multipole power
spectrum $P_\ell(\vk)$ to trace the underlying power on a range of
scales $\vk'$ and multipoles $\ell'$.

We produce a practical algorithm for evaluating
$S_{\ell,\ell'}(\vk,\vk')$ by employing the addition theorem for
spherical harmonics
\begin{equation}
L_\ell(\hk.\hx) = \frac{4\pi}{2\ell + 1} \sum_{m=-\ell}^\ell 
Y_{\ell,m}^*(\hk) \, Y_{\ell,m}(\hx) ,
\end{equation}
such that
\begin{equation}
S_{\ell,\ell'}(\vk,\vk') = \frac{(4\pi)^2}{(2\ell + 1)(2\ell'+1)}
\sum_{m=-\ell}^\ell Y^*_{\ell,m}(\hk) \sum_{m'=-\ell'}^{\ell'}
Y_{\ell',m'}(\hk') \, \tilde{S}_{\ell,m,\ell',m'}(\vk-\vk') ,
\end{equation}
where
\begin{equation}
\tilde{S}_{\ell,m,\ell',m'}(\vk) = \frac{1}{V} \int d^3\vx \, n_w(\vx)
\, Y_{\ell,m}(\hx) \, Y^*_{\ell',m'}(\hx) \, e^{i\vk.\vx} .
\label{eqslm}
\end{equation}
Hence we obtain the final result
\begin{equation}
  \langle \hP_\ell(k) \rangle = \frac{1}{I} \int
  \frac{d\Omega_k}{4\pi} \sum_{m=-\ell}^\ell Y_{\ell,m}(\hk)
  \sum_{\ell'} \frac{(4\pi)^2}{2\ell'+1} \sum_{m'=-\ell'}^{\ell'}
  \left[ \frac{V}{(2\pi)^3} \int d^3\vk' \, P_{\ell'}(k') \,
    Y^*_{\ell',m'}(\hk') \, \tilde{n}_w(\vk-\vk') \,
    \tilde{S}^*_{\ell,m,\ell',m'}(\vk-\vk') \right] .
\label{eqpkconvharm}
\end{equation}
This relation expresses how the expectation value $\langle \hP_\ell(k)
\rangle$ mixes the underlying multipole power spectra $P_{\ell'}(k')$
in a harmonic sum of convolutions involving the survey window function
and mixing terms $\tilde{S}_{\ell,m,\ell',m'}$.  The evaluation of the
model power spectrum multipoles is hence carried out purely in Fourier
space as intended\footnote{The alternative formulation discussed below
  and in Appendix \ref{secapp} uses Hankel transforms between Fourier
  and configuration space.  These may be performed efficiently using
  FFTlog \citep{Hamilton00}, but do require careful convergence study
  of integration ranges.}: the formulation of Equation
\ref{eqpkconvharm} can be evaluated for each $\ell$ using a framework
of FFTs by computing $\tilde{S}_{\ell,m,\ell',m'}(\vk)$ using Equation
\ref{eqslm}, performing a convolution of $P_{\ell'} \, Y^*_{\ell',m'}$
and $\tilde{n}_w \, \tilde{S}^*_{\ell,m,\ell',m'}$ after distributing
$P_{\ell'}(k')$ on a 3D Fourier grid, summing over multipoles
$(m,\ell',m')$, and averaging the final power spectrum in bins of $k =
|\vk|$.  In principle the sum over $\ell'$ is infinite, but in
practice it converges within 2 or 3 terms (which we verify in Section
\ref{sec6dfgsmod}).  If there are $N_\ell$ power spectrum multipoles
$P_{\ell'}(k)$ in the model, the sum over $(m,\ell',m')$ involves
$(2\ell+1) \times N_\ell \times (2N_\ell - 1)$ terms, where each term
requires the evaluation of four real-to-complex FFTs.\footnote{After
  the result for each $(m,\ell',m')$ is added to the sum for $\langle
  \hP_\ell(k) \rangle$, this memory may be freed.}

We check the special case of convolution with a constant window
function $n(\vx) = n_0$ and $w(\vx) = 1$, for which $\tilde{n}_w(\vk)
= n_0 \, \tilde{\delta}_D(\vk)$ where $\tilde{\delta}_D(\vk)$ is the
Dirac $\delta$-function in Fourier space, normalized such that
$[V/(2\pi)^3] \int \delta^3\vk \, \tilde{f}(\vk) \,
\tilde{\delta}_D(\vk - \vk_0) = \tilde{f}(\vk_0)$.  Substituting this
relation in Equation \ref{eqpkl4} we obtain
\begin{equation}
  \langle \hP_\ell(k) \rangle = \frac{(2\ell + 1) \, n_0}{I} \int
  \frac{d\Omega_k}{4\pi} \, \sum_{\ell'} P_{\ell'}(k) \,
  S_{\ell,\ell'}(\vk,\vk) ,
\end{equation}
where, using Equation \ref{eqsl},
\begin{equation}
S_{\ell,\ell'}(\vk,\vk) = \frac{n_0}{V} \int d^3\vx \, L_\ell(\hk.\hx)
\, L_{\ell'}(\hk.\hx) = \frac{n_0 \, \delta_{\ell \ell'}}{2\ell + 1} ,
\end{equation}
by orthogonality of the Legendre polynomials.  Since $I = n_0^2$, we
then find $\langle \hP_\ell(k) \rangle = P_\ell(k)$, as expected in
the absence of a survey window function.

Equation \ref{eqpkconvharm} is an alternative formulation of the
mathematics presented by \citet{Wilson17} and \citet{Beutler17}, which
we derive in Appendix \ref{secapp} for completeness.  These authors
show that the convolved multipole power spectra may be written in the
form
\begin{equation}
\langle \hP_\ell(k) \rangle = 4\pi \, i^\ell \int ds \, s^2 \,
j_\ell(ks) \, \hxi_\ell(s) ,
\label{eqpkconvxi}
\end{equation}
where $j_\ell$ are the spherical Bessel functions and
\begin{equation}
\hxi_\ell(s) = (2\ell + 1) \sum_{\ell'} \xi_{\ell'}(s) \sum_{\ell''}
A^{\ell''}_{\ell,\ell'} \, \frac{W_{\ell''}^2(s)}{2\ell'' + 1} ,
\label{eqxiconv}
\end{equation}
is written in terms of the model correlation function multipoles
$\xi_\ell$, coefficients $A^{\ell''}_{\ell,\ell'}$ listed in Equation
\ref{eqal}, and window function multipoles
\begin{equation}
W_\ell^2(s) = \frac{2\ell+1}{I} \int \frac{d\Omega_s}{4\pi} \,
\frac{1}{V} \int d^3\vx \, n_w(\vx) \, n_w(\vx+\vs) \, L_\ell(\hx.\hs)
.
\label{eqwl}
\end{equation}
\citet{Wilson17} and \citet{Beutler17} recommend evaluating $W_\ell^2$
in bins of separation $s$ of width $\Delta s$ using a pair count over
a random catalogue, where each pair $(\vx,\vx')$ is weighted in
proportion to $w(\vx) \, w(\vx') \, L_\ell(\vx.\vs)/s^2 \, \Delta
s$.\footnote{We note that the random catalogue must be replicated
  periodically in such an analysis to avoid spurious edge effects, as
  can be seen by considering the evaluation of $W_0^2(s)$ for a
  uniform window function within a cuboid, which should reproduce
  $W_0^2 = 1$ for all $s$ such that there is no convolution.}
Alternatively, using the FFT-based formalism of our study to avoid the
pair count, we can evaluate the window function multipoles by writing
Equation \ref{eqwl} in the form
\begin{equation}
W_\ell^2(s) = \frac{2\ell+1}{I} \int \frac{d\Omega_s}{4\pi} \,
\frac{1}{V} \int d^3\vx \, n_w(\vx) \, \frac{V}{(2\pi)^3} \int d^3\vk \,
\tilde{n}_w^*(\vk) \, e^{i\vk.\vx} \, e^{i\vk.\vs} \, L_\ell(\hx.\hs) .
\end{equation}
We produce a practical algorithm for evaluating this expression by
employing the plane-wave expansion
\begin{equation}
e^{i\vk.\vs} = \sum_\ell i^\ell \, (2\ell+1) \, j_\ell(ks) \,
L_\ell(\hk.\hs) ,
\label{eqplane}
\end{equation}
to derive the new relation
\begin{equation}
  W_\ell^2(s) = \frac{4\pi \, i^\ell}{I} \sum_{m=-\ell}^\ell
  \frac{V}{(2\pi)^3} \int d^3\vk \, \tilde{n}_w^*(\vk) \, j_\ell(ks)
  \, Y_{\ell,m}(\hk) \, \frac{1}{V} \int d^3\vx \, n_w(\vx) \,
  Y_{\ell,m}^*(\hx) \, e^{i\vk.\vx} .
\label{eqwin}
\end{equation}
This expression may be evaluated for given $\ell$ and $s$ by computing
the FFT of $n_w(\vx) \, Y_{\ell,m}^*(\hx)$, and summing the product of
functions over $\vk$-space, for each $m$.  We note that this is the
generalization of Equation 20 of \citet{Wilson17}, which applies in
the flat-sky approximation.  The two formulations are identical for
$\ell = 0$.

Multipole power spectrum estimates may be corrected to satisfy the
integral constraint condition, which can be evaluated as described by
Section A2 of \citet{Beutler17}.  The correction is typically
negligible on the scales of interest.

\subsection{Cross-power spectrum multipoles with overlapping tracers}
\label{secmult}

The preceding formulation may be extended to the analysis of
spatially-overlapping galaxy populations via their cross-power
spectrum \citep[see also,][]{Smith09,Blake13}.  The generalization of
Equation \ref{eqest} to the estimation of the cross-power spectrum
multipoles $P_{\ell,c}(k)$ of two tracers with number densities
$n_1(\vx)$ and $n_2(\vx)$, with respective weight functions $w_1(\vx)$
and $w_2(\vx)$, is given by
\begin{equation}
  \hP_{\ell,c}(k) = \frac{2\ell + 1}{I_c} \int \frac{d\Omega_k}{4\pi}
  \frac{1}{V} \int d^3\vx \int d^3\vx' \, \frac{1}{2} \left[
    \delta_{1,w}(\vx) \, \delta_{2,w}(\vx') + \delta_{2,w}(\vx) \,
    \delta_{1,w}(\vx') \right] L_\ell(\hk.\hx') \, e^{i\vk.(\vx-\vx')}
  ,
\label{eqestcross}
\end{equation}
in terms of the weighted overdensities
\begin{equation}
\delta_{i,w}(\vx) = w_i(\vx) \left[ n_i(\vx) - \langle n_i(\vx)
  \rangle \right] ,
\end{equation}
where $i = \{ 1,2 \}$ denotes the tracer, and the normalization of the
cross-power spectrum
\begin{equation}
I_c = \frac{1}{V} \int d^3\vx \, w_1(\vx) \, w_2(\vx) \, \langle
n_1(\vx) \rangle \, \langle n_2(\vx) \rangle .
\end{equation}
We note that Equation \ref{eqestcross} is constructed to be symmetric
under interchange of the indices 1 and 2, and that there is no shot
noise contribution to the cross-power spectrum.  We estimate the
cross-power spectrum multipoles using a lightly-modified version of
the method of \citet{Bianchi15}.  We can evaluate the expectation
value of Equation \ref{eqestcross} using
\begin{equation}
\langle \delta_{1,w}(\vx) \, \delta_{2,w}(\vx') \rangle = w_1(\vx) \,
w_2(\vx') \, \langle n_1(\vx) \rangle \, \langle n_2(\vx') \rangle \,
\xi_{12}(\vx,\vx') ,
\end{equation}
where $\xi_{12}$ is the cross-correlation function of the tracers,
which is related to their cross-power spectrum in the same manner as
Equation \ref{eqxi}.  Proceeding with a similar derivation as in
Section \ref{secconv} we find
\begin{equation}
  \langle \hP_{\ell,c}(k) \rangle = \frac{1}{I_c} \int
  \frac{d\Omega_k}{4\pi} \sum_{m=-\ell}^\ell Y_{\ell,m}(\hk)
  \sum_{\ell'} \frac{(4\pi)^2}{2\ell'+1} \sum_{m'=-\ell'}^{\ell'}
  \left[ \left( P_{\ell',c} \, Y^*_{\ell',m'} \right) \star
    \frac{1}{2} \left( \tilde{n}_{2,w} \, \tilde{S}^{1
      *}_{\ell,m,\ell',m'} + \tilde{n}_{1,w} \, \tilde{S}^{2
      *}_{\ell,m,\ell',m'} \right) \right] ,
\end{equation}
where $n_{i,w}(\vx) = w_i(\vx) \, \langle n_i(\vx) \rangle$,
$\tilde{S}^i_{\ell,m,\ell',m'}$ is given by Equation \ref{eqslm} with
$n_w = n_{i,w}$, and $\star$ denotes convolution.

\section{Covariance of the power spectrum multipoles}
\label{seccov}

In this Section we formulate analytical expressions for the covariance
of the power spectrum multipole estimates described in Section
\ref{secest}, between different multipoles and scales.  \citet{FKP}
derive the covariance of the estimated galaxy power spectrum
\begin{equation}
\langle \delta \hP(k) \, \delta \hP(k') \rangle = \left\langle \left(
\hP(k) - \langle \hP(k) \rangle \right) \left( \hP(k') - \langle
\hP(k') \rangle \right) \right\rangle = \langle \hP(k) \, \hP(k')
\rangle - \langle \hP(k) \rangle \langle \hP(k') \rangle ,
\end{equation}
in the approximation that the galaxy overdensity is a Gaussian random
field sampled by Poisson statistics.  We use similar methods to
determine analogous expressions for the joint covariance of the
multipole power spectra, extending the calculations of
\citet{Taruya10}, \citet{Grieb16} and \citet{Hand17} to account for
window function and curved-sky effects (in the local plane-parallel
approximation).

We note that the covariance may be impacted by various non-Gaussian
effects, such as higher-order correlations (trispectrum terms)
\citep[e.g.][]{OConnell16,Howlett17} or non-Poisson sampling of the
density field by galaxies
\citep[e.g.][]{Seljak09,Baldauf13,Ginzburg17}.  However, Gaussian
approximations may serve a valuable purpose on large scales, or when
simulations are not available to calibrate these effects sufficiently.

The covariance between power spectrum multipoles $(\ell,\ell')$
averaged in spherical shells around wavenumbers $(k,k')$ may be
determined as
\begin{equation}
\langle \delta\hP_\ell(k) \, \delta\hP_{\ell'}(k') \rangle = \int
\frac{d\Omega_k}{4\pi} \int \frac{d\Omega_{k'}}{4\pi} \langle
\delta\hP_\ell(\vk) \, \delta\hP_{\ell'}(\vk') \rangle ,
\end{equation}
where, using the estimator of Equation \ref{eqest},
\begin{equation}
\begin{split}
  \langle \delta\hP_\ell(\vk) \, \delta\hP_{\ell'}(\vk') \rangle =
  \frac{(2\ell+1) \, (2\ell'+1)}{V^2 \, I^2} &\int d^3\vx_1 \int
  d^3\vx_2 \int d^3\vx_3 \int d^3\vx_4 \, e^{i\vk.(\vx_1-\vx_2)} \,
  e^{i\vk'.(\vx_3-\vx_4)} \, L_\ell(\hk.\hx_2) \,
  L_{\ell'}(\hk'.\hx_4) \\
&\left[ \langle \delta_w(\vx_1) \, \delta_w(\vx_2) \, \delta_w(\vx_3)
    \, \delta_w(\vx_4) \rangle - \langle \delta_w(\vx_1) \,
    \delta_w(\vx_2) \rangle \langle \delta_w(\vx_3) \, \delta_w(\vx_4)
    \rangle \right] .
\end{split}
\label{eqcov1}
\end{equation}
We apply the Gaussian approximation using Wick's theorem for a
Gaussian random field
\begin{equation}
\langle \delta_1 \, \delta_2 \, \delta_3 \, \delta_4 \rangle = \langle
\delta_1 \, \delta_2 \rangle \langle \delta_3 \, \delta_4 \rangle +
\langle \delta_1 \, \delta_3 \rangle \langle \delta_2 \, \delta_4
\rangle + \langle \delta_1 \, \delta_4 \rangle \langle \delta_2 \,
\delta_3 \rangle ,
\end{equation}
which (omitting some algebra) allows us to write the covariance in the
form
\begin{equation}
\langle \delta\hP_\ell(\vk) \, \delta\hP_{\ell'}(\vk') \rangle =
\frac{(2\ell+1) \, (2\ell'+1)}{I^2} \left[ C_{\ell,\ell'}(\vk,\vk') \,
  C_{0,0}(\vk,\vk')^* + C_{\ell,0}(\vk,\vk') \,
  C_{0,\ell'}(\vk,\vk')^* \right] ,
\end{equation}
where
\begin{equation}
C_{\ell,\ell'}(\vk,\vk') = \frac{1}{V} \int d^3\vx \int d^3\vx' \,
\langle \delta_w(\vx) \, \delta_w(\vx') \rangle \, e^{i(\vk.\vx -
  \vk'.\vx')} \, L_\ell(\hk.\hx) \, L_{\ell'}(\hk'.\hx') .
\end{equation}
We evaluate this expression by substituting in Equations \ref{eqdel}
and \ref{eqxi} with the additional approximation that the covariance
between two scales $\vk$ and $\vk'$ may be determined by the
clustering power at those scales, rather than by all modes (i.e., for
the purposes of evaluating the covariance, neglecting convolution).
This approximation is necessary to allow the covariance to be
expressed as a single integral over space, which may be evaluated by
FFTs.  We find
\begin{equation}
C_{\ell,\ell'}(\vk,\vk') \approx \frac{1}{V} \int d^3\vx \, w^2(\vx)
\, \left[ n^2(\vx) \, P_{\rm eff}(\vk,\vk',\vx) + n(\vx) \right]
e^{i(\vk-\vk').\vx} \, L_\ell(\hk.\hx) \, L_{\ell'}(\hk'.\hx) ,
\label{eqcov2}
\end{equation}
where, in order to preserve the symmetry of the covariance
$C_{\ell,\ell}(\vk,\vk') = C_{\ell,\ell}(\vk',\vk)$, we use an
effective power spectrum
\begin{equation}
  P_{\rm eff}(\vk,\vk',\vx) = \frac{1}{2} \sum_{\ell''} \left[
    P_{\ell''}(k) \, L_{\ell''}(\hk.\hx) + P_{\ell''}(k') \,
    L_{\ell''}(\hk'.\hx) \right] ,
\end{equation}
in place of Equation \ref{eqpk}.  In physical terms, Equation
\ref{eqcov2} represents a combination of sample variance and shot
noise contributions, respectively $C^{SV}$ and $C^{SN}$, corresponding
to the two terms inside the square bracket.  In practice we restrict
the sum over $\ell''$ to the monopole and quadrupole.

We illustrate our method for evaluating Equation \ref{eqcov2} using the
first term, $P_{\rm eff} = \sum_{\ell''} P_{\ell''}(k) \,
L_{\ell''}(\hk.\hx)$.  Its contribution to the sample variance
covariance is
\begin{equation}
C^{SV}_{\ell,\ell'}(\vk,\vk') = \sum_{\ell''} P_{\ell''}(k) \,
\frac{1}{V} \int d^3\vx \, n_w^2(\vx) \, L_{\ell''}(\hk.\hx) \,
L_\ell(\hk.\hx) \, L_{\ell'}(\hk'.\hx) \, e^{i(\vk-\vk').\vx} .
\label{eqcov3}
\end{equation}
Using Equation \ref{eqal} this can be written in the form
\begin{equation}
C^{SV}_{\ell,\ell'}(\vk,\vk') = \sum_{\ell''} P_{\ell''}(k)
\sum_{\ell'''} A_{\ell,\ell''}^{\ell'''}
\frac{(4\pi)^2}{(2\ell'''+1)(2\ell'+1)} \sum_{m'''=-\ell'''}^{\ell'''}
\sum_{m'=-\ell'}^{\ell'} Y^*_{\ell''',m'''}(\hk) \, Y_{\ell',m'}(\hk')
\, \tilde{Q}_{\ell''',m''',l',m'}(\vk-\vk') ,
\label{eqcov4}
\end{equation}
where
\begin{equation}
\tilde{Q}_{\ell,m,\ell',m'}(\vk) = \frac{1}{V} \int d^3\vx \, n_w^2(\vx)
\, Y_{\ell,m}(\hx) \, Y^*_{\ell',m'}(\hx) \, e^{i\vk.\vx} .
\label{eqqlm}
\end{equation}
This covariance contribution $C^{SV}$ quantifies how the underlying
power spectrum drives sample variance, which is correlated across
scales by the window function $\tilde{Q}$.\footnote{We note that
  $\tilde{Q}$ is the Fourier transform of the product of $n_w(\vx) \,
  Y_{\ell,m}(\hx)$ and $n_w(\vx) \, Y^*_{\ell',m'}(\hx)$, which is the
  convolution of the $(\ell,m)$ and $(\ell',m')$ moments of $n_w$ in
  Fourier space.}  Likewise for the shot noise contribution:
\begin{equation}
C^{SN}_{\ell,\ell'}(\vk,\vk') = \frac{1}{V} \int d^3\vx \, n_w(\vx) \,
L_\ell(\hk.\hx) \, L_{\ell'}(\hk'.\hx) \, e^{i(\vk-\vk').\vx} =
\frac{(4\pi)^2}{(2\ell+1)(2\ell'+1)} \sum_{m=-\ell}^\ell
\sum_{m'=-\ell'}^{\ell'} Y^*_{\ell,m}(\hk) \, Y_{\ell',m'}(\hk') \,
\tilde{S}_{\ell,m,l',m'}(\vk-\vk') ,
\end{equation}
where $\tilde{S}_{\ell,m,l',m'}$ is defined by Equation \ref{eqslm}.
This hence provides an FFT-based framework for evaluating the
covariance in the Gaussian approximation, by first computing
$\tilde{S}_{\ell,m,\ell',m'}(\vk)$ and
$\tilde{Q}_{\ell,m,\ell',m'}(\vk)$ using Equations \ref{eqslm} and
\ref{eqqlm} and then performing sums over spherical harmonic
coefficients.  Our development has hence reduced the dimensionality of
the covariance calculation from a multiple integral over four 3D
spaces in Equation \ref{eqcov1}, to a harmonic sum of terms sampled
within a single $\vk$-space.

It is useful to consider some special cases.  For a constant window
function $n(\vx) = n_0$ and $w(\vx) = 1$, the appearance of the
position-dependent Legendre polynomials inside Equation \ref{eqcov3}
implies that there are still correlations between multipole power
spectra $P_\ell(\vk)$ with different $\ell$ and $\vk$.  This is unlike
the case of a global flat-sky approximation relative to an axis
$\hx_0$, for which the Legendre polynomials in Equation \ref{eqcov3}
take the form $L_\ell(\hk.\hx_0)$ and the integral over $\vx$ becomes
a Dirac $\delta$-function, removing the $\vk$-correlations.  Another
useful special case is an isotropic window function, for which
Equation \ref{eqqlm} can be written
\begin{equation}
\tilde{Q}_{\ell,m,\ell',m'}(0) = \frac{1}{V} \int dr \, r^2 \,
n_w^2(r) \int d\Omega \, Y_{\ell,m}(\hx) \, Y^*_{\ell',m'}(\hx) =
\frac{1}{V} \int dr \, r^2 \, n_w^2(r) \, \delta_{\ell,\ell'}
\delta_{m,m'} ,
\end{equation}
using the orthonormality of the spherical harmonics.  Using this
result in Equation \ref{eqcov4} with $\vk' = \vk$, we find
\begin{equation}
  C^{SV}_{\ell,\ell'}(\vk,\vk) = \frac{1}{V} \int dr \, r^2 \,
  n_w^2(r) \sum_{\ell''} P_{\ell''}(k) \, \left( \begin{matrix} \ell &
    \ell'' & \ell' \\ 0 & 0 & 0 \end{matrix} \right)
\end{equation}
where $\left( \begin{matrix} \ell & \ell'' & \ell' \\ 0 & 0 &
  0 \end{matrix} \right)$ is a Wigner 3j-symbol.

For a general window function we have not found any further
simplications, but the terms are straight-forward to evaluate
numerically.  In practice we carry out the required average by
randomly sub-sampling modes $\vk$ and $\vk'$ in spherical shells
around wavenumbers $k$ and $k'$, separately treating the case $\vk' =
\vk$.\footnote{We note that, in the lowest-$k$ shells, a lack of FFT
  modes in the bins may imprint systematics when evaluating these sums
  \citep{Wilson17}.}  We randomly choose (a maximum of) 100 modes in
each bin to perform this calculation, checking that our results are
not sensitive to this choice.  For example, to determine the
contribution of Equation \ref{eqcov4} to the first covariance term we
compute
\begin{equation}
    \int \frac{d\Omega_k}{4\pi} \int \frac{d\Omega_{k'}}{4\pi} \,
    C^{SV}_{\ell,\ell'}(\vk,\vk') \, C_{0,0}(\vk,\vk')^* =
    \sum_{\ell''} P_{\ell''}(k) \sum_{\ell'''}
    A_{\ell,\ell''}^{\ell'''} \frac{(4\pi)^2}{(2\ell'''+1)(2\ell'+1)}
    \sum_{m'''} \sum_{m'} \left[ P_0(k') \, \overline{QQ}(k,k') +
      \overline{QS}(k,k') \right] ,
\end{equation}
where
\begin{equation}
\begin{split}
\overline{QQ}(k,k') &= \overline{Y^*_{\ell,m}(\hk) \,
  Y_{\ell',m'}(\hk') \, \tilde{Q}_{\ell,m,\ell',m'}(\vk-\vk') \,
  \tilde{Q}^*_{0,0,0,0}(\vk-\vk')} \\
\overline{QS}(k,k') &= \overline{Y^*_{\ell,m}(\hk) \,
  Y_{\ell',m'}(\hk') \, \tilde{Q}_{\ell,m,\ell',m'}(\vk-\vk') \,
  \tilde{S}^*_{0,0,0,0}(\vk-\vk')} ,
\end{split}
\end{equation}
where the overbar denotes an average over a sub-sample of modes.  We
derive similar expressions for the remaining terms.

In a multi-tracer analysis, we can also develop the remaining
expressions for the covariances of the auto- and cross-power spectrum
multipoles in the Gaussian approximation, i.e., $\langle
\hP_{\ell,1}(k) \, \hP_{\ell',2}(k') \rangle$, $\langle
\hP_{\ell,1}(k) \, \hP_{\ell',c}(k') \rangle$ and $\langle
\hP_{\ell,c}(k) \, \hP_{\ell',c}(k') \rangle$.  For example, the first
of these covariances can be evaluated by averaging in $k$-space shells
the quantity
\begin{equation}
\langle \delta\hP_{\ell,1}(\vk) \, \delta\hP_{\ell',2}(\vk') \rangle =
\frac{(2\ell+1) \, (2\ell'+1)}{I_1 \, I_2} \left[
  C^{12}_{\ell,\ell'}(\vk,\vk') \, C^{12}_{0,0}(\vk,\vk')^* +
  C^{12}_{\ell,0}(\vk,\vk') \, C^{12}_{0,\ell'}(\vk,\vk')^* \right] ,
\end{equation}
where
\begin{equation}
  C^{12}_{\ell,\ell'}(\vk,\vk') \approx \frac{1}{V} \int d^3\vx \,
  n_{1,w}(\vx) \, n_{2,w}(\vx) \, P_c(\vk,\vk',\vx) \, e^{i(\vk-\vk').\vx}
  \, L_\ell(\hk.\hx) \, L_{\ell'}(\hk'.\hx) .
\end{equation}
Using the same methods as above, the equivalent of Equation
\ref{eqcov4} is
\begin{equation}
C^{12}_{\ell,\ell'}(\vk,\vk') = \sum_{\ell''} P_{c,\ell''}(k)
\sum_{\ell'''} A_{\ell,\ell''}^{\ell'''}
\frac{(4\pi)^2}{(2\ell'''+1)(2\ell'+1)} \sum_{m'''} \sum_{m'}
Y^*_{\ell''',m'''}(\hk) \, Y_{\ell',m'}(\hk') \,
\tilde{Q}^{12}_{\ell''',m''',l',m'}(\vk-\vk') ,
\end{equation}
where
\begin{equation}
\tilde{Q}^{12}_{\ell,m,\ell',m'}(\vk) = \frac{1}{V} \int d^3\vx \,
n_{1,w}(\vx) \, n_{2,w}(\vx) \, Y_{\ell,m}(\hx) \, Y^*_{\ell',m'}(\hx)
\, e^{i\vk.\vx} .
\end{equation}
As before, we numerically compute the average by randomly sub-sampling
modes $\vk$ and $\vk'$ in spherical shells around wavenumbers $k$ and
$k'$:
\begin{equation}
\int \frac{d\Omega_k}{4\pi} \int \frac{d\Omega_{k'}}{4\pi} \,
C^{12}_{\ell,\ell'}(\vk,\vk') \, C^{12}_{0,0}(\vk,\vk')^* =
\sum_{\ell''} P_{c,\ell''}(k) \sum_{\ell'''} A_{\ell,\ell''}^{\ell'''}
\frac{(4\pi)^2}{(2\ell'''+1)(2\ell'+1)} \sum_{m'''} \sum_{m'}
P_{c,0}(k') \, \overline{QQ^{12}}(k,k') ,
\end{equation}
where
\begin{equation}
\overline{QQ^{12}}(k,k') = \overline{Y^*_{\ell,m}(\hk) \,
  Y_{\ell',m'}(\hk') \, \tilde{Q}^{12}_{\ell,m,\ell',m'}(\vk-\vk') \,
  \tilde{Q}^{12 *}_{0,0,0,0}(\vk-\vk')} .
\end{equation}
Using this approach we can deduce expressions for the full joint
covariance of $[P_{\ell,1}(k), P_{\ell,2}(k), P_{\ell,c}(k)]$ in the
Gaussian approximation, where $P_{\ell,i}(k) = [P_{0,i}(k),
  P_{2,i}(k), P_{4,i}(k), ...]$.  An example application of this
methodology to a mock catalogue of the DESI Bright Galaxy Survey is
presented by DESI collaboration et al.\ (in prep.).

\begin{figure}
\includegraphics[width=\columnwidth]{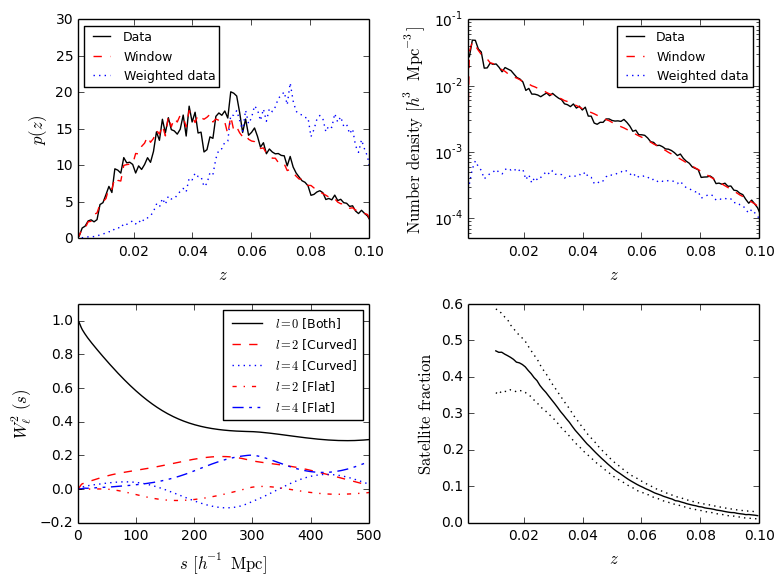}
\caption{The upper left-hand and right-hand panels display,
  respectively, the redshift probability distribution and number
  density distribution averaged in redshift slices of the 6dFGS data
  sample (black solid line), the window function (red dashed line) and
  the data including the weights (blue dotted line).  The lower
  left-hand panel plots the first three multipoles $W_\ell^2(s)$ of
  the window function in the range $s < 500 \, h^{-1}$ Mpc using
  curved-sky and flat-sky formulations, illustrating the importance of
  curved-sky effects for the $\ell > 0$ terms.  The lower right-hand
  panel illustrates the redshift dependence of the average satellite
  fraction in the mocks (with the standard deviation across mocks
  indicated by the dotted lines).}
\label{fignz}
\end{figure}

\section{Application to the 6-degree Field Galaxy Survey}
\label{sec6dfgs}

\subsection{Data and mocks}

The 6-degree Field Galaxy Survey \citep[6dFGS,][]{Jones09,Springob14}
is a southern-hemisphere survey of $\sim 10^5$ galaxy redshifts and
$\sim 10^4$ peculiar velocities carried out by the UK Schmidt
Telescope between 2001 and 2006.  Covering $17{,}000$ deg$^2$ of sky
south of declination $0^\circ$ and located more than $10^\circ$ from
the Galactic plane, and mapping galaxies with median redshift $0.053$,
the 6dFGS remains one of the widest-angle studies of large-scale
structure in the local Universe.\footnote{The 2MASS Redshift Survey
  \citep[2MRS,][]{Huchra12} is an all-sky, somewhat shallower,
  spectroscopic galaxy survey.}  The dataset has been subject to a
number of previous cosmological analyses, including measurement and
fitting of the baryon acoustic peak \citep{Beutler11}, RSD fits to the
2D correlation function \citep{Beutler12}, comparison studies of RSD
around galaxies and voids \citep{Achitouv17}, and analysis of the
cross-correlation of peculiar velocities and densities
\citep{Adams17}.  The growth rate of structure has also been measured
using the statistics of the 6dFGS peculiar velocity field
\citep{Johnson14,Huterer17,Adams17}.  As far as we are aware, there is
no existing Fourier-space analysis of RSD using 6dFGS data.

In this Section we analyze the power spectrum multipoles of a subset
of the 6dFGS redshift catalogue, forming a sample selected with
magnitude $K < 12.9$ in sky regions with redshift completeness greater
than $60\%$, which is the sample used in the 6dFGS baryon acoustic
peak study of \citet{Beutler11}.\footnote{We note that the 6dFGS RSD
  study of \citet{Beutler12} used a slightly brighter faint magnitude
  limit $K < 12.75$ in order to render the completeness corrections
  negligible, although we expect the results to be comparable.}  We
restrict our analysis to galaxies in the redshift range $0.001 < z <
0.1$, given that the 6dFGS number density is very low at higher
redshifts, producing a final sample of $70{,}467$ objects across
$15{,}419$ deg$^2$.  We map galaxies onto a 3D co-moving grid using a
fiducial flat cosmological model with matter density $\Omega_m = 0.3$
(matching the fiducial cosmology used to generate the mock catalogues,
described below).  When computing FFTs we enclose the hemispheric
catalogue by a cuboid with dimensions $(586,586,293) \, h^{-1}$ Mpc
and grid size $(128,128,64)$, such that the Nyquist sampling frequency
in each dimension is $k_{\rm Nyq} \approx 0.7 \, h$ Mpc$^{-1}$ (where
we fit models to the range $k < 0.2 \, h$ Mpc$^{-1}$).  We correct for
this gridding effect in our power spectrum estimation.

We compute the window function $\langle n(\vx) \rangle$ of the dataset
following \citet{Jones09} and \citet{Beutler11}, including the angular
spectroscopic completeness and the variation of the redshift
distribution with magnitude (hence local completeness).  Galaxies are
assigned an optimal weight depending on their position, combining the
effects of sample variance and Poisson noise following \citet{FKP}:
\begin{equation}
w(\vx) = \frac{1}{1 + \langle n(\vx) \rangle \, P_0} ,
\label{eqwei}
\end{equation}
where $P_0$ is a characteristic power spectrum amplitude.  Following
\citet{Beutler12}, we take $P_0 = 1{,}600 \, h^{-3}$ Mpc$^3$ (finding
that our results are not overly sensitive to this choice).  The upper
panels of Figure \ref{fignz} illustrate the redshift probability
distribution and number density variation of the 6dFGS data and window
function, including the effect of the weights assigned by Equation
\ref{eqwei}, which up-weight higher redshift (i.e., lower number
density) data compared to lower redshift data, increasing the weighted
mean redshift of the sample from $0.049$ to $0.067$.  Defining the
effective redshift of the power spectrum measurement as an
optimally-weighted sum over the selection function,
\begin{equation}
z_{\rm eff} = \frac{\sum_{\vx} z \, W(\vx)}{\sum_{\vx} W(\vx)}
\end{equation}
where $W(\vx) = \left( \frac{ \langle n(\vx) \rangle P_0 }{ 1 +
  \langle n(\vx) \rangle P_0 } \right)^2$, we find $z_{\rm eff} =
0.060$, which we take as the redshift of our growth rate measurement.

The lower left-hand panel of Figure \ref{fignz} illustrates the first
three window function multipoles $W_\ell^2(s)$, computed using the
FFT-based method of Equation \ref{eqwin} in the range $s < 500 \,
h^{-1}$ Mpc.  These window function multipoles are used in Section
\ref{sec6dfgsmod} to test the convolution framework based on
correlation function multipoles, compared to the purely Fourier-space
method.  We also determine the window function multipoles in the
flat-sky approximation (i.e., using Equation 20 of
\citealt{Wilson17}).  The evaluations are identical for $\ell = 0$,
but differ significantly for $\ell > 0$, illustrating the importance
of curved-sky effects for wide-area surveys.

We also use a set of $N_{\rm mock} = 600$ mock catalogues produced as
part of the 6dFGS baryon acoustic peak reconstruction analysis
\citep{Carter18}.  These mock catalogues are generated by populating a
set of dark matter halos produced by fast N-body techniques with a
Halo Occupation Distribution of central and satellite galaxies,
designed to match the redshift distribution and projected clustering
of the 6dFGS sample; we refer the reader to \citet{Carter18} for more
details.  The fiducial cosmological model used for the initial
conditions of the mocks was $\Omega_m = 0.3$, baryon density $\Omega_b
= 0.0478$, Hubble parameter $h = 0.68$, clustering amplitude $\sigma_8
= 0.82$ and spectral index $n_s = 0.96$.  The fiducial normalized
growth rate at $z = 0$ is then $f \sigma_8 = 0.423$.  We use the mocks
to test the recovery of the input growth rate using our procedure, and
to compare the Gaussian covariance with that deduced from the ensemble
of mocks.  The lower right-hand panel of Figure \ref{fignz} displays
the mean and standard deviation of the satellite fraction as a
function of redshift across the 600 mocks; we will comment further on
the effects of satellites in Section \ref{sec6dfgserr}.

\subsection{Power spectrum multipole models}
\label{sec6dfgsmod}

For the purposes of this study we describe the dependence of the
redshift-space galaxy power spectrum, $P(k,\mu)$, on the cosine of the
angle to the line of sight, $\mu$, in terms of a simple 3-parameter
model \citep{Peacock94,Cole95,Hatton98} combining the large-scale
Kaiser effect \citep{Kaiser87} imprinted by the growth rate $f$,
exponential damping from random pairwise velocities with dispersion
$\sigma_v$, and a linear galaxy bias $b$:
\begin{equation}
  P(k,\mu) = \frac{(b + f \mu^2)^2 \, P_m(k)}{1 + \left( k \mu
    \sigma_v/H_0 \right)^2} ,
\label{eqmod}
\end{equation}
where $P_m(k)$ is the model matter power spectrum, which we compute
using a recent version of the CAMB software package \citep{Lewis00},
including the non-linear `halofit' correction
\citep{Smith03,Takahashi12}, for the fiducial cosmological parameters
listed above.  Given that $P_m(k) \propto \sigma_8^2$ on large scales,
it is convenient for RSD analyses to adopt the parameter set $(f
\sigma_8, b \sigma_8)$.  The inclusion of the Hubble parameter $H_0 =
100 \, h$ km s$^{-1}$ Mpc$^{-1}$ in Equation \ref{eqmod} implies that
the pairwise velocity dispersion is measured in units of km s$^{-1}$
at $z = 0$.  The multipole power spectra may be computed from Equation
\ref{eqmod} as
\begin{equation}
P_\ell(k) = \frac{2\ell+1}{2} \int_{-1}^1 d\mu \, P(k,\mu) \,
L_\ell(\mu) .
\end{equation}
Three multipoles $(\ell = 0,2,4)$ are sufficient to describe the model
in linear theory $(\sigma_v = 0)$; in practice the hexadecapole
contains extremely low signal-to-noise for current datasets such that
analyses, including ours, focus on solely the monopole and quadrupole.

Many enhancements to the simple model of Equation \ref{eqmod} have
been proposed, incorporating more accurate corrections for non-linear
effects in the evolution of densities, velocities and galaxy bias
\citep[e.g.,][]{Scoccimarro04,Taruya10,Jennings11,Reid11}.  As we are
focussing on testing the estimation, convolution and covariance of the
clustering statistics, rather than performing the best possible
measurement of the growth rate, the simple model of Equation
\ref{eqmod} suits our purpose and we do not consider more
sophisticated theoretical treatments.

\begin{figure}
\includegraphics[width=\columnwidth]{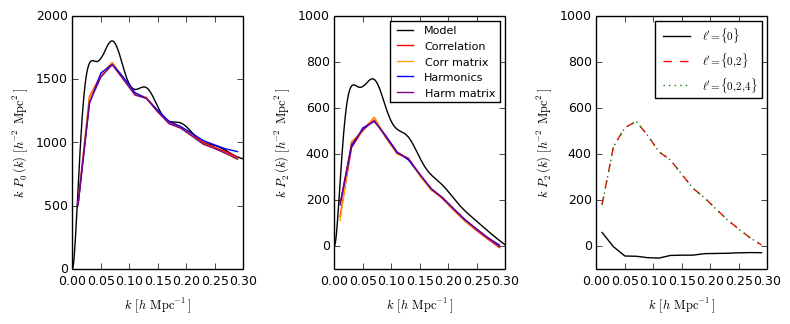}
\caption{A consistency check of different determinations of the
  convolved monopole (left-hand panel) and quadrupole (middle panel)
  of a fiducial model (black line) with the survey window function, in
  a series of Fourier bins of width $\Delta k = 0.02 \, h$ Mpc$^{-1}$.
  Evaluations are shown using the correlation function and window
  function multipoles (red line), a convolution matrix derived by
  applying this method to a set of unit vectors as described in the
  text (orange line), the method using spherical harmonics and FFTs
  (blue line), and a corresponding convolution matrix (purple line).
  These different methods produce convolved power spectra in good
  agreement with each other.  The right-hand panel displays a test of
  the number of multipoles $\ell'$ in Equation \ref{eqpkconvharm}
  required for the evaluation of the convolved quadrupole to converge.
  We see that excellent convergence is obtained using 2 terms.  The
  power spectra are scaled by a factor of $k$ for clarity of
  presentation.}
\label{figconv}
\end{figure}

Figure \ref{figconv} illustrates the convolution of model power
spectrum multipoles with the 6dFGS window function, computed using the
two approaches described in Section \ref{secconv} in bins of width
$\Delta k = 0.02 \, h$ Mpc$^{-1}$ in the range $k < 0.3 \, h$
Mpc$^{-1}$.  We use a fiducial parameter choice $f \sigma_8 = 0.423$,
$\sigma_v = 300$ km s$^{-1}$ and $b \sigma_8 = 1.19$, which is close
to the best-fitting values for the 6dFGS data.  We note the good
agreement between the determination of the convolution using spherical
harmonics and FFTs based on Equation \ref{eqpkconvharm}, and the
evaluation via the correlation function and window function
multipoles, using Equations \ref{eqpkconvxi}, \ref{eqxiconv} and
\ref{eqwin}.  Small differences occur due to the sparse distribution
of the grid of Fourier wavevectors $\vk$ within each spherical shell
in $k$-space.  Figure \ref{figconv} also verifies that the convolved
quadrupole converges when Equation \ref{eqpkconvharm} is evaluated
using the first two terms in $\ell'$.

Rather than re-compute a full convolution for each different RSD
model, we determine a convolution matrix $M_{ij}$ in the Fourier bins
$i$, such that the convolved multipole power spectrum $\hP$, which is
the concatenation of $(\hP_0, \hP_2, \hP_4)$, may be determined from
the unconvolved power spectrum $P = (P_0, P_2, P_4)$ as $\hP_i =
\sum_j M_{ij} \, P_j$.  We evaluate the convolution matrix by applying
the full convolution (Equation \ref{eqpkconvharm}) to a set of unit
model vectors $P_j$, generated such that $P_\ell(k) = 1$ for the
relevant multipole within a single bin $k_{{\rm min},j} < k < k_{{\rm
    max},j}$, and $P_\ell(k) = 0$ for the remaining $k$-values and
multipoles.  We tested that this approach produced a negligible change
in results compared to implementing the full convolution whilst being
orders of magnitude faster to evaluate; these results are also
displayed in Figure \ref{figconv}.

\subsection{Power spectrum multipole measurements and errors}
\label{sec6dfgserr}

\begin{figure}
\includegraphics[width=\columnwidth]{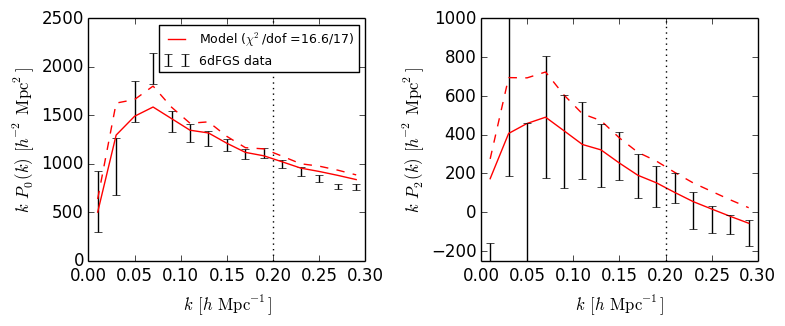}
\caption{The monopole (left-hand panel) and quadrupole (right-hand
  panel) power spectra of the $z < 0.1$ 6dFGS $K$-band galaxy sample,
  applying FKP weighting with characteristic amplitude $P_0 = 1{,}600
  \, h^{-3}$ Mpc$^3$.  The dashed and solid lines show, respectively,
  the unconvolved and convolved best-fitting 3-parameter RSD model fit
  to the monopole and quadrupole in the wavenumber range $k < 0.2 \,
  h$ Mpc$^{-1}$ (indicated by the vertical dotted lines), with the
  corresponding $\chi^2$ value indicated in the plot.  Convolution
  with a window function reduces the amplitude of a model power
  spectrum model, because a selection function always contains a
  smaller effective volume than the cube in which it is embedded, such
  that the convolved power is carried by a higher number density of
  Fourier modes \citep{Peacock91}.  The errors are determined using
  the Gaussian approximation to the covariance matrix, generated from
  this best-fitting model.  The power spectra are scaled by a factor
  of $k$ for clarity of presentation.}
\label{figpk}
\end{figure}

The monopole and quadrupole power spectra of the 6dFGS large-scale
structure dataset, measured in 15 Fourier bins of width $\Delta k =
0.02 \, h$ Mpc$^{-1}$ in the range $k < 0.3 \, h$ Mpc$^{-1}$, are
displayed in Figure \ref{figpk} together with the best-fitting model.
We postpone a discussion of the best-fitting model parameters to the
next section, and focus here on the errors in the measured statistics.
We note the upward sample variance fluctuation in the monopole at the
scale corresponding to the baryon acoustic peak ($k \approx 0.07 \, h$
Mpc$^{-1}$), which may contribute toward the strong signature of the
acoustic peak in the 6dFGS correlation function reported by
\citet{Beutler11}, which was found by that study to lie in the upper
quartile of statistical realizations.  The hexadecapole power spectrum
does not contain any useful signal and we do not consider it further.

\begin{figure}
\includegraphics[width=\columnwidth]{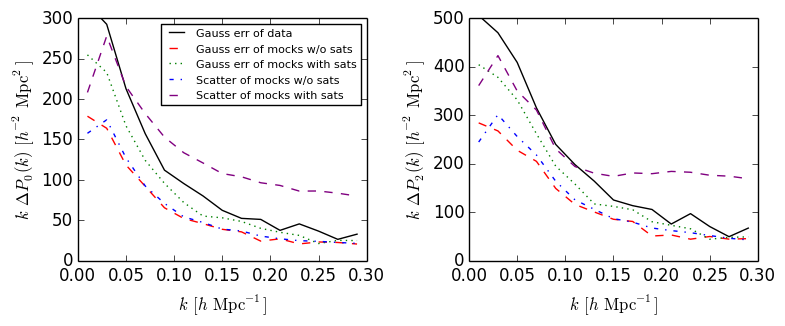}
\caption{Comparison of the errors in the monopole (left-hand panel)
  and quadrupole (right-hand panel) power spectra of the 6dFGS data
  sample and corresponding mock catalogues, including and excluding
  satellites.  The different error determinations apply either the
  Gaussian approximation to the covariance, or measure the standard
  deviation of the power spectrum multipole estimates across the mock
  catalogues.  The inclusion of satellites increases the standard
  deviation of the mock power spectra above the Gaussian estimate; the
  amplitude variation in the Gaussian covariance determinations at low
  $k$ is driven by the different fiducial bias factors in each case.
  The power spectra are scaled by a factor of $k$ for clarity of
  presentation.}
\label{figerr}
\end{figure}

Figure \ref{figerr} compares the standard deviation of the monopole
and quadrupole power spectrum (i.e., the square root of the diagonal
values of the covariance matrix, $\sqrt{C_{ii}}$) for the 6dFGS data
and mocks, estimated using different techniques.  These determinations
use either the Gaussian covariance of Section \ref{seccov}, or are
estimated from the ensemble of mock catalogues as
\begin{equation}
  C_{ij} = \frac{1}{N_{\rm mock}-1} \left[ \sum_k \left( P_{i,k} -
    \overline{P_i} \right) \left( P_{j,k} - \overline{P_j} \right)
    \right] = \frac{N_{\rm mock}}{N_{\rm mock}-1} \left( \overline{P_i
    P_j} - \overline{P_i} \; \overline{P_j} \right) ,
\end{equation}
where $P_i$ is the concatenation of the monopole and quadrupole in a
combined data vector of 30 entries, $P_{i,k}$ is the measurement of
these values in the $k$-th mock, and $\overline{P_i} = (1/N_{\rm
  mock}) \sum_{k=1}^{N_{\rm mock}} P_{i,k}$ is the average value
across the ensemble of mocks.  The Gaussian covariance evaluation uses
the window function $\langle n(\vx) \rangle$ and requires a choice of
fiducial multipole power spectra for each sample.  We compute the
fiducial power spectra from Equation \ref{eqmod} with $f \sigma_8 =
0.423$, $\sigma_v = 300$ km s$^{-1}$ and $b \sigma_8 = (1.19, 1.04,
0.84)$ for the data, mocks including satellites and mocks excluding
satellites, respectively, which are close to the best-fitting values
in each case.  The inclusion of satellites in the mock increases the
fiducial bias factor.

It is interesting to compare mock catalogues including and excluding
satellite galaxies.  The inclusion of satellites, which are located
preferentially in high-mass halos, enhances the clustering power by
up-weighting massive halos, and causes the Poisson sampling model
described by Equation \ref{eqdel} to break down
\citep{Baldauf13,Ginzburg17}.  For central-only galaxy samples, the
Gaussian error agrees well with the dispersion across the mock
catalogues, as illustrated by the comparison of the red and blue lines
in Figure \ref{figerr}.  The inclusion of satellites complicates the
picture, increasing the standard deviation of the mock power spectra
above the Gaussian estimate.  Since the sample variance contribution
to the covariance scales as $(b \sigma_8)^2$ on large scales, the
different fiducial bias values drive the amplitude variations in the
Gaussian covariance at low $k$.

\begin{figure}
\includegraphics[width=\columnwidth]{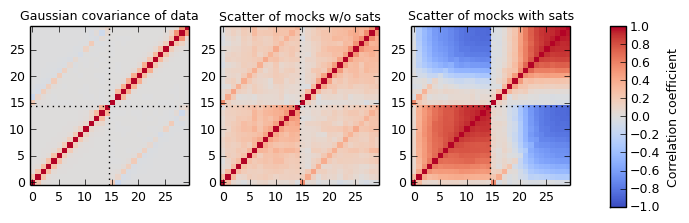}
\caption{The correlation matrix of the monopole and quadrupole,
  derived from the covariance matrix $C_{ij}$ as $C_{ij}/\sqrt{C_{ii}
    \, C_{jj}}$.  The 30 bins correspond to the concatenation $(P_0,
  P_2)$ for measurements up to $k = 0.3 \, h$ Mpc$^{-1}$ in bins of
  width $\Delta k = 0.02 \, h$ Mpc$^{-1}$.  Three cases are shown,
  from left-to-right: the 6dFGS data sample covariance determined in
  the Gaussian approximation, and the covariance deduced from the
  ensemble of mock catalogues excluding satellites and including
  satellites.}
\label{figcov}
\end{figure}

Figure \ref{figcov} provides further insight by illustrating the
correlation matrix, which quantifies the amplitude of the off-diagonal
covariance compared to the diagonal elements as $C_{ij}/\sqrt{C_{ii}
  \, C_{jj}}$, corresponding to various cases.  The left-hand panel
displays the covariance of the 6dFGS data in the Gaussian
approximation, evaluated using the methodology of Section
\ref{seccov}.  The central panel is the covariance evaluated across
the ensemble of mock catalogues, excluding satellites.  The
correlation structure of the principal diagonals and first
off-diagonal of the two matrices is similar.  The mock-based matrix
contains somewhat greater correlation than the Gaussian approximation
between bins that are widely separated in $k$, typically with a
cross-correlation coefficient less than $0.2$, but at a somewhat
higher level for monopole covariance at $k > 0.2 \, h$ Mpc$^{-1}$
(noting that we do not fit to these scales).  The right-hand panel
displays the mock covariance including satellite galaxies, featuring a
significant enhancement in the off-diagonal covariance with increasing
$k$.  Evidently, satellites in the 6dFGS HOD mock (which constitute a
significant fraction of the sample, as illustrated by Figure
\ref{fignz}) have a strong impact in correlating different scales of
the power spectrum monopole and quadrupole, and anti-correlating the
monopole and quadrupole.  This effect may arise because the
trispectrum component, which drives significant off-diagonal
covariance in Fourier space \citep{Howlett17}, is boosted by the
significant satellite fraction in these mocks, and the effect is
extended across a wide range of scales by the significant width of the
$k$-space window function of this small-volume survey.  In the next
Section we will explore the impact of these differences on the RSD
parameter fits for the 6dFGS data and mocks.

\subsection{RSD parameter fits}

\begin{figure}
\includegraphics[width=\columnwidth]{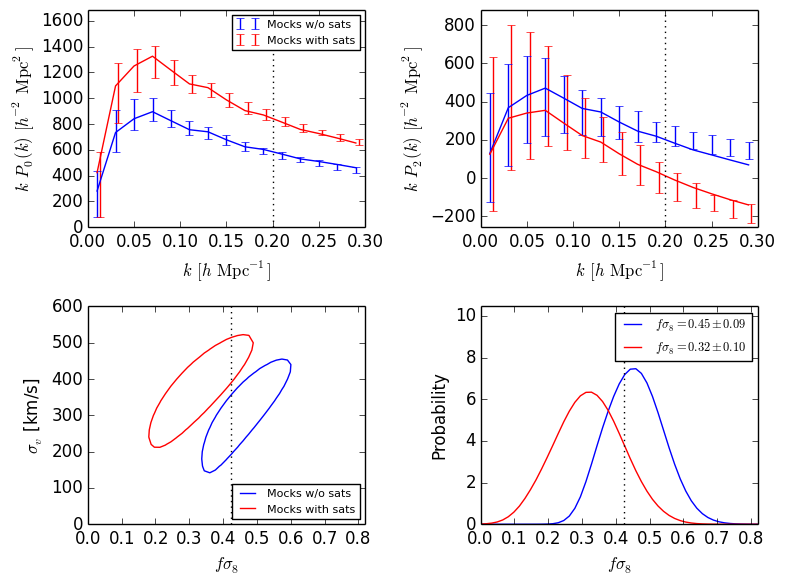}
\caption{The power spectrum multipole measurements and RSD parameter
  fits obtained from the 6dFGS mock catalogues, considering cases
  excluding and including satellites (displayed in blue and red,
  respectively, and offset along the $x$-axis for clarity).  The
  upper-left and upper-right panels display the mock mean monopole and
  quadrupole power spectra, with errors determined using the Gaussian
  approximation to the covariance.  The solid lines in these panels
  are the best-fitting models in each case, with the vertical dotted
  line indicating the maximum fitted scale ($k = 0.2 \, h$
  Mpc$^{-1}$).  The lower-left panel shows the $68\%$ confidence
  region of the 2D posterior probability distribution for the
  normalized growth rate $f \sigma_8$ and velocity dispersion
  $\sigma_v$, marginalized over the galaxy bias $b \sigma_8$, and the
  right-hand panel shows the marginalized probability distribution for
  the growth rate.  The growth rate ranges quoted in the right-hand
  panel are $68\%$ confidence intervals.  The vertical dotted line in
  the lower panels indicates the growth rate expected in the fiducial
  $\Omega_m = 0.3$ model.  The inclusion of satellites causes a
  systematic shift in the best-fitting growth rate, owing to the
  breakdown of the exponential damping model.  However, this shift is
  smaller than the statistical error in the measurement.}
\label{figrsdmock}
\end{figure}

We fit the RSD power spectrum model of Equation \ref{eqmod} to the
monopole and quadrupole of the 6dFGS data and mocks over the range $k
< 0.2 \, h$ Mpc$^{-1}$, varying the 3 parameters $(f \sigma_8,
\sigma_v, b \sigma_8)$.  We first consider the results of fitting to
the mock catalogues, both excluding and including satellites.  The
mean power spectrum monopole and quadrupole, averaged across the 600
mocks, is displayed in the upper panels of Figure \ref{figrsdmock},
with the error (as appropriate for a single mock) computed using the
Gaussian approximation to the covariance.  The increased low-$k$
errors in the power spectra for the mocks with satellites is driven by
the increased effective galaxy bias factor, recalling that the sample
variance contribution to the covariance scales as $(b \sigma_8)^2$ on
large scales.  We overplot the best-fitting RSD models, which provide
a good description of the measurements over the fitted range.

The lower panels of Figure \ref{figrsdmock} illustrate the posterior
probabilities of the RSD parameters.  The marginalized measurements of
the normalized growth rate are $f \sigma_8 = (0.45 \pm 0.09)$ and
$(0.32 \pm 0.10)$ for the sample excluding and including satellites,
respectively, compared to the fiducial value of $0.42$.  Satellites
create systematic error in the model of Equation \ref{eqmod}, where a
simple exponential damping scenario is unable to represent their
effect in detail.  We conclude that, at the level of precision
afforded by the 6dFGS sample, the model of Equation \ref{eqmod}
provides an acceptable description of the clustering pattern over the
range $k < 0.2 \, h$ Mpc$^{-1}$, although it would suffer from
systematic modelling errors given significantly larger datasets,
especially due to the presence of satellites.  For comparison, we note
that an RSD fit to $k < 0.1 \, h$ Mpc$^{-1}$ produces marginalized
measurements $f \sigma_8 = (0.43 \pm 0.11)$ and $(0.37 \pm 0.13)$
excluding and including satellites, in closer agreement with the
fiducial value of $0.42$, and a fit to $k < 0.3 \, h$ Mpc$^{-1}$
produces fits with significantly enhanced systematic errors,
respectively $f \sigma_8 = (0.50 \pm 0.06)$ and $(0.31 \pm 0.06)$.

\begin{figure}
\includegraphics[width=\columnwidth]{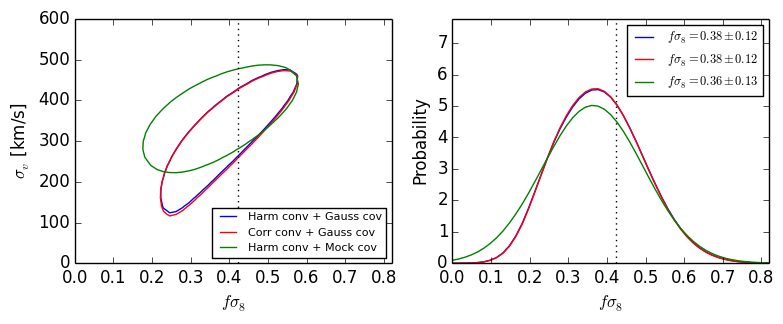}
\caption{RSD parameter fits to the 6dFGS galaxy sample.  The left-hand
  panel displays the $68\%$ confidence region of the 2D posterior
  probability distribution for the normalized growth rate $f \sigma_8$
  and velocity dispersion $\sigma_v$, marginalized over the galaxy
  bias $b \sigma_8$, and the right-hand panel shows the marginalized
  probability distribution for the growth rate.  Results are displayed
  for three cases: using spherical harmonics convolution and the
  covariance matrix derived using the Gaussian approximation (`Harm
  conv + Gauss cov', blue contours); correlation function convolution
  with the same covariance (`Corr conv + Gauss cov', red contours) and
  spherical harmonics convolution with the covariance matrix derived
  from mock catalogues (`Harm conv + Mock cov', green contours).  The
  growth rate ranges quoted in the right-hand panel are $68\%$
  confidence intervals.  The vertical dotted line indicates the growth
  rate expected in a fiducial $\Omega_m = 0.3$ model.  The analyses
  using different convolution and covariance approaches produce
  consistent results.}
\label{figparfit}
\end{figure}

We now consider the corresponding RSD parameter fits to the 6dFGS data
sample, also using the fitting range $k < 0.2 \, h$ Mpc$^{-1}$.  The
best-fitting unconvolved and convolved models using the Gaussian
covariance are overplotted in Figure \ref{figpk}, and Figure
\ref{figparfit} displays the $68\%$ confidence region of the joint
posterior probability distribution for $(f \sigma_8, \sigma_v)$
(left-hand panel), and the marginalized probability distribution for
the normalized growth rate (right-hand panel).  We consider various
cases.  Our fiducial analysis, using our new convolution treatment
described in Section \ref{secconv} and the Gaussian covariance matrix,
produces marginalized parameter fits $f \sigma_8 = 0.38 \pm 0.12$,
$\sigma_v = 290 \pm 120$ km s$^{-1}$ and $b \sigma_8 = 1.19 \pm 0.03$
(quoting $68\%$ confidence regions), with minimum $\chi^2 = 16.6$ for
17 degrees-of-freedom.  Replacing the convolution scheme with a method
using window function and correlation function multipoles produces
almost identical results.  Using the covariance matrix deduced from
mock catalogues including satellites produces consistent best-fitting
parameters (the growth rate fit is $f \sigma_8 = 0.36 \pm 0.13$), but
the minimum $\chi^2$ increases to $34.3$, driven by the off-diagonal
terms of the covariance matrix shown in Figure
\ref{figcov}.\footnote{When performing fits using the inverse
  covariance matrix deduced from the ensemble of mock catalogues, we
  include the corrections discussed by \citet{Hartlap07}, although
  they are very small in this case.}

Our growth rate fitted to the 6dFGS power spectrum multipoles agrees
well with previous analyses of the 6dFGS redshift and peculiar
velocity samples.  \citet{Beutler12} and \citet{Achitouv17} both
report a measurement $f \sigma_8 = 0.42 \pm 0.06$ obtained by fitting
RSD models to the 2D galaxy correlation function, and similar results
are obtained when fitting to the 2-point clustering of peculiar
velocities and density-velocity correlations
\citep{Johnson14,Huterer17,Adams17}.  The error in our measured growth
rate is roughly a factor of 2 higher than these correlation function
studies.  We compared our fitted error in the growth rate to that
forecast by a standard Fisher matrix calculation
\citep[e.g.,][]{White09,Abramo12,Blake13}.  Assuming an idealized
survey with the area and redshift distribution of our 6dFGS sample and
neglecting covariance between Fourier modes, we forecast an error
$\sigma(f \, \sigma_8) = 0.10$ for a fit to the wavenumber range $k <
0.2 \, h$ Mpc$^{-1}$.  The error in our measurement is only $20\%$
larger than this forecast, suggesting that it is realistic and that
the correlation function fits are accessing information on smaller
scales.  Improved Fourier-space modelling would be required to extend
the power spectrum fits to these scales.

\section{Summary}
\label{secsummary}

The correlated peculiar velocities of galaxies, generated by the
gravitational physics of the clustered density field, modulate
redshift-space galaxy clustering with respect to the local
line of sight.  These directional anisotropies may be described in
terms of a multipole expansion of the clustering pattern.  The
changing line-of-sight direction across a galaxy survey complicates
the analysis of clustering multipoles, particularly in Fourier space,
where the survey volume is embedded within an enclosing cuboid to
allow the application of efficient FFT techniques.

In this study we have explored three extensions of power spectrum
multipole calculations.  First, we derived an alternative formulation
of the convolution of the power spectrum multipoles with a survey
window function in a curved sky, in the local plane-parallel
approximation.  This expression is evaluated purely in Fourier space
using FFTs, and produces very similar results to approaches using the
correlation function and window function multipoles, whilst avoiding
the need for mapping the statistics into configuration space.  Second,
we developed expressions for the joint covariance of power spectrum
multipoles within a Gaussian approximation, including window function
and curved-sky effects.  Finally, we generalized these results to
include the cross-power spectrum multipoles of overlapping galaxy
tracers.  Accompanying python code is available at
\url{https://github.com/cblakeastro/multipoles}.

We applied our framework to conduct the first Fourier-space clustering
analysis of the 6-degree Field Galaxy Survey.  By fitting a simple RSD
model to the monopole and quadrupole in the range $k < 0.2 \, h$
Mpc$^{-1}$, we determined a best-fitting growth rate $f \sigma_8 =
0.38 \pm 0.12$ at effective redshift $z = 0.06$, in agreement with
previous 6dFGS analyses.  The statistical error of our measurement
agrees with a Fisher matrix error forecast, although is twice as large
as previous correlation function analyses, which access smaller
scales.  We verified our fitting pipeline using a series of mock
catalogues, demonstrating the impact of satellite galaxies on
covariance properties.  We note that techniques such as cylindrical
exclusion \citep{Okumura17} can reduce the impact of satellites on
clustering analyses.

Future galaxy redshift surveys such as the Taipan Galaxy Survey
\citep{daCunha17}, the Dark Energy Spectroscopic Instrument
\citep{DESI} and the 4MOST Cosmology Redshift Survey \citep{4MOST}
will allow the clustering of the local and distant Universe to be
quantified with increasing accuracy, and cosmological models to be
precisely tested.  Power spectrum multipoles will continue to provide
a valuable connection between observations and theory.

\section*{Acknowledgements}

We are grateful to the anonymous referee for thoroughly reviewing the
paper.  We also thank Florian Beutler, Michael Wilson and Cullan
Howlett for extremely valuable comments on a draft of this paper.

JK is supported by MUIR PRIN 2015 "Cosmology and Fundamental Physics:
illuminating the Dark Universe with Euclid" and Agenzia Spaziale
Italiana agreement ASI/INAF/I/023/12/0.

The 6dF Galaxy Survey was made possible by contributions from many
individuals towards the instrument, the survey and its science.  We
particularly thank Matthew Colless, Heath Jones, Will Saunders, Fred
Watson, Quentin Parker, Mike Read, Lachlan Campbell, Chris Springob,
Christina Magoulas, John Lucey, Jeremy Mould and Tom Jarrett, as well
as the dedicated staff of the Australian Astronomical Observatory and
other members of the 6dFGS team over the years.

Part of this work was performed on the swinSTAR supercomputer at
Swinburne University of Technology.  We have used {\tt matplotlib}
\citep{Hunter07} for the generation of scientific plots, and this
research also made use of {\tt astropy}, a community-developed core
Python package for Astronomy \citep{Astropy13}.

\bibliographystyle{mnras}
\bibliography{6dfgs_multipoles}

\appendix

\section{Derivation of convolution using correlation function multipoles}
\label{secapp}

For completeness we derive in this Appendix the evaluation of Equation
\ref{eqpkl3} using a multipole expansion of the correlation function
\citep{Wilson17,Beutler17}, $\xi(\vx,\vx') = \sum_{\ell'}
\xi_{\ell'}(s) \, L_{\ell'}(\hs.\hx')$ where $\vs = \vx' - \vx$, such
that
\begin{equation}
  \langle \hP_\ell(k) \rangle = \frac{2\ell + 1}{I} \int
  \frac{d\Omega_k}{4\pi} \, \frac{1}{V} \int d^3\vx \, n_w(\vx) \int
  d^3\vs \, n_w(\vx+\vs) \, e^{i\vk.\vs} \, L_\ell(\hk.\hx)
  \sum_{\ell'} \xi_{\ell'}(s) \, L_{\ell'}(\hs.\hx) .
\end{equation}
We note that the model correlation function multipoles $\xi_\ell$ may
be determined from the power spectrum multipoles using the inverse
transform of Equation \ref{eqpkconvxi}:
\begin{equation}
  \xi_\ell(s) = \frac{(-i)^\ell}{2\pi^2} \int dk \, k^2 \, j_\ell(ks)
  \, P_\ell(k) .
\label{eqxil}
\end{equation}
Using the plane wave expansion of Equation \ref{eqplane} we obtain
\begin{equation}
  \langle \hP_\ell(k) \rangle = \frac{2\ell + 1}{I} \frac{1}{V} \int
  d^3\vx \, n_w(\vx) \int d^3\vs \, n_w(\vx+\vs) \sum_{\ell'}
  \xi_{\ell'}(s) \, L_{\ell'}(\hs.\hx) \sum_{\ell''} i^{\ell''} \,
  (2\ell'' + 1) \, j_{\ell''}(ks) \int \frac{d\Omega_k}{4\pi} \,
  L_\ell(\hk.\hx) \, L_{\ell''}(\hk.\hs) .
\end{equation}
Using the result $\int \frac{d\Omega_k}{4\pi} \, L_\ell(\hk.\hx) \,
L_{\ell'}(\hk.\hs) = \frac{1}{2\ell+1} \, L_\ell(\hs.\hx) \,
\delta_{\ell,\ell'}$, we can simplify this expression to the form
\begin{equation}
  \langle \hP_\ell(k) \rangle = \frac{2\ell + 1}{I} \frac{1}{V} \int
  d^3\vx \, n_w(\vx) \int d^3\vs \, n_w(\vx+\vs) \sum_{\ell'}
  \xi_{\ell'}(s) \, L_{\ell'}(\hs.\hx) \, i^\ell \, j_\ell(ks) \,
  L_\ell(\hs.\hx) .
\end{equation}
It is convenient to write this relation in the form of an inverse
transform to Equation \ref{eqxil},
\begin{equation}
\langle \hP_\ell(k) \rangle = 4\pi \, i^\ell \int ds \, s^2 \,
j_\ell(ks) \, \hxi_\ell(s) ,
\end{equation}
where
\begin{equation}
\hxi_\ell(s) = \frac{2\ell + 1}{I} \, \frac{1}{V} \int d^3\vx \,
n_w(\vx) \int \frac{d\Omega_s}{4\pi} \, n_w(\vx+\vs) \sum_{\ell'}
\xi_{\ell'}(s) \, L_\ell(\hs.\hx) \, L_{\ell'}(\hs.\hx) .
\end{equation}
We can conveniently evaluate this expression using the Legendre
expansion of a product of Legendre polynomials \citep{Adams78}:
\begin{equation}
  L_\ell(x) \, L_{\ell'}(x) = \sum_{\ell''=|\ell-\ell'|}^{\ell+\ell'} \left( \begin{matrix} \ell
    & \ell' & \ell'' \\ 0 & 0 & 0 \end{matrix} \right)^2 \, (2\ell'' +
  1) \, L_{\ell''}(x) = \sum_{\ell''} A^{\ell''}_{\ell,\ell'} \,
  L_{\ell''}(x) ,
\label{eqal}
\end{equation}
where $\left( \begin{matrix} \ell & \ell' & \ell'' \\ 0 & 0 &
  0 \end{matrix} \right)$ is a Wigner 3j-symbol, and the coefficients
$A^{\ell''}_{\ell,\ell'}$ are given by \citet{Bailey33} as
\begin{equation}
  A^{\ell''}_{\ell,\ell'} = \left( \begin{matrix} \ell & \ell' &
    \ell'' \\ 0 & 0 & 0 \end{matrix} \right)^2 \, (2\ell'' + 1) =
  \frac{G_{\ell-p} \, G_p \, G_{\ell'-p}}{G_{\ell+\ell'-p}} \left(
  \frac{2\ell + 2\ell' - 4p + 1}{2\ell + 2\ell' - 2p + 1} \right) ,
\end{equation}
in which $G_p = [1.3...(2p-1)]/[p!]$, $\ell'' = \ell + \ell' - 2p$
and, in the sum over $\ell''$ defined by Equation \ref{eqal}, $p$
ranges from $0$ to ${\rm min}(\ell,\ell')$.  Using this notation we
can write
\begin{equation}
\hxi_\ell(s) = (2\ell + 1) \sum_{\ell'} \xi_{\ell'}(s) \sum_{\ell''}
A^{\ell''}_{\ell,\ell'} \, \frac{W_{\ell''}^2(s)}{2\ell'' + 1} ,
\end{equation}
in terms of the window function multipoles
\begin{equation}
W_\ell^2(s) = \frac{2\ell+1}{I} \int \frac{d\Omega_s}{4\pi} \,
\frac{1}{V} \int d^3\vx \, n_w(\vx) \, n_w(\vx+\vs) \, L_\ell(\hx.\hs)
,
\end{equation}
in agreement with Equation A9 of \citet{Beutler17}.  For example
\begin{equation}
\hxi_0(s) = W_0^2(s) \, \xi_0(s) + \frac{1}{5} \, W_2^2(s) \, \xi_2(s)
+ \frac{1}{9} \, W_4^2(s) \, \xi_4(s) + ...
\end{equation}
since if $\ell=0$, then $p=0$ and $A^{\ell'}_{0,\ell'} = 1$.
Relations for the other ``convolved'' multipoles $\hxi_\ell$, in terms
of $\xi_\ell$, are provided by \citet{Wilson17} and \citet{Beutler17}.

We can use Equation \ref{eqwin} to check a couple of useful special
cases.  First,
\begin{equation}
  W_0^2(0) = \frac{1}{I} \frac{V}{(2\pi)^3} \int d^3\vk \,
  |\tilde{n}_w(\vk)|^2 = \frac{1}{I} \, \frac{1}{V} \int d^3\vx \,
  n_w^2(\vx) = 1 .
\end{equation}
Also, for a constant window function $n(\vx) = n_0$ we find
\begin{equation}
  W_\ell^2(s) = \frac{4\pi \, i^\ell}{n_0^2} \sum_{m=-\ell}^\ell \,
  n_0 \, j_\ell(0) \, Y_{\ell,m}(\hz) \, \frac{1}{V} \int d^3\vx \, n_0
  \, Y_{\ell,m}^*(\hx) = \delta_{\ell,0} ,
\end{equation}
where $\delta_{\ell,0}$ is a Kronecker delta.

\bsp
\label{lastpage}
\end{document}